\documentclass[aps,prc,reprint,superscriptaddress,nopreprintnumbers]{revtex4-1}

\usepackage{graphicx}

\usepackage{todonotes}
\usepackage{tikz}
\usetikzlibrary{calc,intersections,through,backgrounds,patterns,decorations.pathreplacing,shapes,arrows}
\tikzstyle{decision} = [diamond, draw, fill=blue!20, 
    text width=7em, text badly centered, node distance=3cm, inner sep=0pt]
\tikzstyle{block} = [rectangle, draw, fill=blue!20, 
    text width=7em, text centered, rounded corners, minimum height=4em]
\tikzstyle{line} = [draw, -latex']
\tikzstyle{cloud} = [draw, text centered, circle,fill=red!20, node distance=3cm,
    text width=7em, minimum height=5em]

\usepackage[english]{babel}
\usepackage[reqno,sumlimits,intlimits]{amsmath}
\usepackage{tabularx}
\usepackage{array} 
\newcolumntype{M}[1]{>{\centering\arraybackslash}m{#1}} 
\usepackage[caption=false]{subfig}
\usepackage[repeatunits=false,tophrase=-,per=fraction,fraction=nice,textmode,expproduct=cdot,alsoload=hep]{siunitx} 
\newunit{\nuclearNumber}{A}
\newunit{\AGeV}{\nuclearNumber\giga\electronvolt}
\newunit{\ATeV}{\nuclearNumber\tera\electronvolt}
\newunit{\fmc}{\femto\metre\per\clight}
\newunit{\fm}{\femto\metre}
\newunit{\GeVc2}{\giga\electronvolt\per\clight\squared}
\newunit{\GeVc}{\giga\electronvolt\per\clight}
\newunit{\GeVpercubicfm}{\giga\electronvolt\per\cubic\fm}

\newcommand{\PbPb}{Pb\,+\,Pb}

\newcommand{\ProtonProton}{p\,+\,p}

\newcommand{\twotwo}{$2\rightarrow2$}
\newcommand{\twothree}{$2\rightarrow3$}

\newcommand{\twothreetwo}{$2\leftrightarrow3$}

\newcommand{\PYTHIA}{\textsc{Pythia}}
\newcommand{\BAMPS}{\textsc{Bamps}}

\newcommand {\Pb}    {${}^{208}\text{Pb}$}
\newcommand{\vect}[1]{\boldsymbol{\mathbf{#1}}}

\def\l{\left}
\def\r{\right}

\begin{document}
\title{Influence of multiple in-medium scattering processes \\ on the momentum imbalance of reconstructed di-jets}

\author{Florian Senzel}
\email{senzel@th.physik.uni-frankfurt.de}
\author{Oliver Fochler}
\author{Jan Uphoff}
\affiliation{Institut f\"ur Theoretische Physik, Johann Wolfgang 
Goethe-Universit\"at Frankfurt, Max-von-Laue-Str.~1, \\
D-60438 Frankfurt am Main, Germany}
\author{Zhe Xu}
\affiliation{Department of Physics, Tsinghua University, Beijing 100084, China}
\author{Carsten Greiner}
\affiliation{Institut f\"ur Theoretische Physik, Johann Wolfgang 
Goethe-Universit\"at Frankfurt, Max-von-Laue-Str.~1, \\
D-60438 Frankfurt am Main, Germany}

\date{\today}

\begin{abstract}
Experimental data measured in $\sqrt{s}=\SI{2.76}{\tera\electronvolt}$ \PbPb{} collisions at the LHC show a significant enhancement of events with an unbalanced pair of reconstructed jet momenta in comparison with \ProtonProton{} collisions. This enhancement of momentum imbalance is supposed to be caused by the different momentum loss of the initial back-to-back di-partons by scatterings within the created dense medium. For investigating the underlying partonic momentum loss we employ the on-shell transport model \BAMPS{} (\underline{B}oltzmann \underline{A}pproach for \underline{M}ulti-\underline{P}arton \underline{S}cattering) for full heavy-ion collisions, which numerically solves the 3+1D Boltzmann equation based on \twotwo{} as well as inelastic \twothreetwo{} scattering processes, together with \PYTHIA{} initial state conditions for the parton showers. Due to the employed test-particle approach jet reconstruction within \BAMPS{} events is not trivial. We introduce a method that nevertheless allows the 
microscopic simulation of the full evolution of the shower particles, recoiled medium particles, and the underlying bulk medium in one common microscopic framework. With this method it is possible to investigate the role of the medium recoil for the momentum imbalance $A_J$ while using well-established background subtraction algorithms. Due to the available particle information in configuration as well as momentum space within \BAMPS{}, it is additionally possible to reproduce the entire evolution of the reconstructed jets within the medium. With this information we investigate the sensitivity of the jet momentum loss from the difference in the partonic in-medium path lengths.
\end{abstract}

\maketitle

\section{Introduction}
As already proposed in the 1980s by \citeauthor{Bjorken:1982tu} \cite{Bjorken:1982tu}, high transverse momentum ($p_t$) partons created in the initial hard partonic scattering processes of heavy-ion collisions are an excellent probe for investigating the properties of the created hot and dense medium. One of the first evidences for this energy loss of hard probes was the attenuation of di-hadron correlations in heavy-ion collisions at RHIC \cite{Adams2005}. This observation was further supported by measurements of the suppression of inclusive hadronic spectra compared to scaled \ProtonProton{} references at RHIC \cite{Adler:2002xw,Adcox:2001jp} as well as at LHC \cite{Aamodt:2010jd,CMS:2012aa}. Due to the large collision energy of the LHC and thereby increased production cross section of high $p_t$ partons, the separation between these partons as well as their associated shower particles and the underlying background medium becomes even more distinct with the result that energy loss studies in terms of 
reconstructed jets become feasible.

By studying reconstructed jets within heavy-ion collisions, both the ATLAS \cite{Aad:2010bu} and CMS experiments \cite{Chatrchyan:2011sx,Chatrchyan:2012nia} reported the measurement of an enhanced number of events with an asymmetric pair of back-to-back reconstructed jets in comparison to vacuum \ProtonProton{} events. An observable that describes this asymmetry in reconstructed jet momenta is the momentum imbalance
\begin{align}
 A_J \left( p_{t;1}, p_{t;2} \right)= \frac{p_{t;1}-p_{t;2}}{p_{t;1}+p_{t;2}} \, ,
\end{align}
where $p_{t;1}$ ($p_{t;2}$) is the transverse momentum of the leading (subleading) jet---the reconstructed jet with the highest (second highest) transverse momentum per event.

Several theoretical approaches describe the measured momentum imbalance $A_J$: Ref.~\cite{Qin:2010mn,CasalderreySolana:2011rq,He:2011pd} employ analytic calculations based on perturbative quantumchromodynamics (pQCD). Also full Monte-Carlo event generators simulating the jet energy loss within hydrodynamic background media \cite{Young:2011qx,Renk:2012cx,Zapp:2013zya} as well as partonic transport models using a static \cite{ColemanSmith:2011rw,ColemanSmith:2012vr} as well as an expanding partonic medium \cite{Ma:2013pha} are available. Since many of the models depend on hydrodynamics for the underlying bulk medium evolution, these models struggle with a microscopic description of the medium and therefore a straight forward investigation of the effect of recoiled medium particles on the reconstructed jets. Within these models, the recoiled medium particle are supposed to be lost into the medium and therefore are counted as jet momentum loss. However, this is not a priori justified, since in nature these 
recoiled particles will also evolve within the medium and thereby may be recovered within the jets. This restoring of jet momenta is neglected within the mentioned models with the exception of ref.~\cite{Zapp:2013zya}, where the effect of recoiled medium particles is considered by sampling the medium recoil locally based on thermal distributions.

In the following we present our method for calculating the momentum imbalance $A_J$ of reconstructed jets in the framework of the microscopic transport model \BAMPS{} \cite{Xu:2004mz}, which considers the full 3+1-dimensional expansion of the partonic medium. This is achieved by solving the relativistic Boltzmann equation based on pQCD cross sections in an improved Gunion-Bertsch approximation while explicitly taking the running coupling on a microscopic level into account. These pQCD interactions lead to a realistic suppression of hadrons in terms of the nuclear modification factor $R_{AA}$ and at the same time a significant amount of elliptic flow $v_2$ during the partonic phase \cite{Uphoff:2014cba}. Among the benefits of \BAMPS{} is the equal treatment of both the shower and the bulk medium within the same approach. In contrast to other models, this allows the investigation of the role of further in-medium scatterings of the recoiled medium particles on a microscopical level. 

The actual reconstruction of jets out of the parton showers simulated by \BAMPS{} is done by the anti-$k_t$ algorithm as implemented in the package \textsc{Fastjet} \cite{Cacciari:2011ma} through out this paper. For more information about the reconstruction of jets both in \ProtonProton{}- and heavy-ion events we refer to ref.~\cite{Salam:2009jx,Cacciari:2010te}. 

The present paper is organized as follows. After a short review of the partonic transport model \BAMPS{} in Sec.~\ref{sec:bamps} we discuss in Sec.~\ref{sec:vacuum} how to model appropriately initial conditions while considering the momentum imbalance of reconstructed jets in ``vacuum'' \ProtonProton{} events in which no medium creation is expected. In Sec.~\ref{sec:simStrategy} we present the method to calculate the in-medium evolution of parton shower  within \BAMPS{} and the strategy when considering further in-medium scattering processes of recoiled medium partons. Finally, we show in Sec.~\ref{sec:results} our results on the momentum imbalance $A_J$ and study the influence of multiple in-medium scattering processes of the recoil partons. Furthermore, we have a closer look on the jet momentum loss in comparison to the initial hard parton and its underlying path length dependence.

\section{Partonic transport model BAMPS}\label{sec:bamps}
\BAMPS{} is a full 3+1D transport model that aims to handle the collective propagation of medium particles as well as energy loss phenomena of high $p_t$ particles of heavy-ion collisions within a common framework. To this end it considers massless on-shell partons, whose evolution is described by the relativistic Boltzmann equation,
\begin{align}
p^{\mu} \partial_{\mu} f(\vect{x}, t) = \mathcal{C}_{22} + \mathcal{C}_{2 \leftrightarrow 3} + ... \, ,
\end{align}
which is solved numerically based on leading order pQCD matrix elements utilizing a stochastic collision algorithm and employing a test-particle ansatz \cite{Xu:2004mz}. This test-particle ansatz consists of a scaling of the number of particles by a factor $N_{\text{test}}$ (the number of test-particles per physical particles), while at the same time the cross sections and thereby the scattering probabilities within \BAMPS{} are decreased by $N_{\text{test}}$. Thus in total the physical interaction rate per particle or the mean free path, respectively, is conserved but the overall attainable statistics of the scattering processes is significantly enhanced.

Included are both elastic \twotwo{} scattering processes, like e.g. $g\,g \rightarrow g\,g$, and inelastic \twothreetwo{} interactions, like e.g. $g\,g \leftrightarrow g\,g\,g$, based on the improved Gunion-Bertsch (GB) approximation \cite{Fochler:2013epa} which cures problems of the ``original Gunion-Bertsch'' matrix element \cite{Gunion:1981qs} in the forward and backward rapidity region of the emitted gluon. Within the GB approximation the improved GB matrix element for the process $X\rightarrow Y + g$ factorizes in the binary matrix element for $X\rightarrow Y$ and a radiative factor $P_g$ \cite{Fochler:2013epa}
\begin{align}
\label{me_gb}
		{\l|\overline{\mathcal{M}}_{X\rightarrow Y + g}\r|}^2 =
	\l|\overline{\mathcal{M}}_{X\rightarrow Y}\r|^2 \, P_g
\end{align}
with
\begin{multline}
\label{gb_pgm_radiation_spectrum}
	P_g = 48 \pi \alpha_s(k_\perp^2) \, (1-\bar{x})^2  \,	\\
  \times\, \l[ \frac{ {\bf k}_\perp}{k_\perp^2} +  \frac{ {\bf q}_\perp - {\bf k}_\perp }{ ({\bf q}_\perp - {\bf k}_\perp)^2+ m_D^2\left(\alpha_s(k_\perp^2)\right)} \r]^2 \ .
\end{multline}
The transverse momentum of the emitted and internal gluons are denoted with ${\bf k}_\perp$ and ${\bf q}_\perp$, respectively. The longitudinal momentum fraction $\bar x$ is related to the rapidity of the emitted gluon via $\bar x = k_\perp e^{|y|}/\sqrt{s}$, where $s$ is the squared centre of mass energy of the interaction. $X\rightarrow Y$ stand for any binary process of light quarks and gluons, while only (Mandelstam) $t$ channel dominated processes (equivalent to $X=Y$) have a finite contribution within the GB approximation. The binary matrix elements are given in the same approximation by
\begin{align}
\label{me_22}
		{\l|\overline{\mathcal{M}}_{X\rightarrow Y}\r|}^2 =
	C_{X\rightarrow Y} \,64  \pi^2 \alpha_s^2(t)	 \frac{s^2}{[t-m_{D}^2(\alpha_s(t))]^2} \ ,
\end{align}
where $C_{X\rightarrow Y}$ is the color factor of the respective process. All internal gluon propagators in Eqs.~\ref{gb_pgm_radiation_spectrum} and \ref{me_22} are screened with the Debye mass $m_D$, which is dynamically computed on the basis of the current parton distribution \cite{Xu:2004mz}.

For modeling the important quantum Landau-Pomeranchuk-Migdal (LPM) effect within our semi-classical transport approach, an effective cutoff function $\theta\left(\lambda-X_{LPM}\,\tau_f\right)$ in the inelastic matrix elements is used where $\lambda$ is the mean free path of the jet particle and $\tau_f$ the formation time of the emitted gluon. For $X_{LPM} = 1$, this effectively allows only independent inelastic scatterings, while $X_{LPM} = 0$ leads to no LPM suppression at all. A more sophisticated implementation of the LPM effect, where also some interference processes occur, should lead to a parameter in the intermediate region $0 < X_{LPM} < 1$. For now, we treat $X_{LPM}$ as a parameter and fix its value based on the $R_{AA}$ of neutral pions at RHIC \cite{Uphoff:2014cba}. Any further divergences resulting from the integration of the pQCD matrix elements are cured by Debye screening. 

In contrast to former \BAMPS{} studies, the running of the coupling $\alpha_s\left( Q^2 \right)$ is explicitly taken into account within this study by setting the scale $Q^2$ to the momentum transfer of the respective channel \cite{Uphoff:2011ad,Uphoff:2012gb,Uphoff:2014cba}. In addition, the coupling in the definition of the Debye mass is also evaluated at the respective scale of the propagators. 

Besides $X_{LPM}$, the scaling factor in the LPM theta function, another parameter of \BAMPS{}, which can be varied in reasonable ranges, is the freeze-out energy density $\epsilon$: Particles that are in regions in which the energy density is below $\epsilon$ do not interact within this time step but stream freely. Throughout out this paper we choose a freeze-out energy density of $\epsilon=\SI{0.6}{\GeVpercubicfm}$, which corresponds via $\epsilon = \frac{48 T^4}{\pi^2}$ to a freeze-out temperature of $T_c=\SI{175}{\MeV}$ in a medium that is in thermal equilibrium and consists of Boltzmann particles \cite{Xu:2008av}.

\begin{figure}
\includegraphics[width=\linewidth]{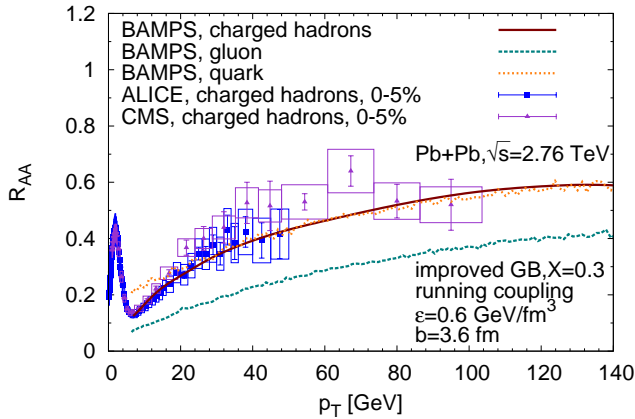}
\caption{(Color online) Nuclear modification factor $R_{AA}$ of gluons, light quarks, and charged hadrons at LHC for a running coupling, the improved Gunion-Bertsch matrix element, and LPM parameter $X_{\rm LPM}=0.3$ together with data of charged hadrons \cite{Abelev:2012hxa}. Figure taken from ref.~\cite{Uphoff:2014cba}.}\label{fig:raa}
\end{figure}

The described interactions based on the improved Gunion-Bertsch matrix element and the microscopically evaluated running coupling show a realistic suppression of single inclusive hadrons not only at RHIC but also at the LHC as shown in fig.~\ref{fig:raa}. As a side remark, in ref.~\cite{Uphoff:2014cba} we showed that the same pQCD interactions do not only account for a realistic suppression of high $p_t$ probes but also for a significant elliptic flow $v_2$ during the partonic stage within the bulk medium regime.

Although the scattering processes of heavy quarks \cite{Uphoff:2010sh,Uphoff:2011ad,Uphoff:2012gb} are already implemented within \BAMPS{}, throughout this work only scatterings of gluons, light quarks and anti-quarks are considered ($N_f=3$). This is motivated by the rare production of initial heavy quarks and therefore minor influence of heavy quark jets in the analysis of the reconstructed jet momentum imbalance.

\section{Simulation strategy for parton showers}\label{sec:simStrategy}
\subsection{Modeling of vacuum parton showers}\label{sec:vacuum}
As it was observed in ref.~\cite{Aad:2010bu,Chatrchyan:2011sx,Chatrchyan:2012nia}, even in \ProtonProton{} collisions, where no medium creation is expected, a momentum asymmetry of the reconstructed jets with the two highest transverse momenta is found. This momentum imbalance is on one hand caused by the probabilistic nature of the independent vacuum splitting processes of the virtual high $p_t$ partons, which are produced in the initial nucleon-nucleon collisions: The evolution in virtuality leads to different out-of-cone splittings of the two back-to-back partons and therefore to an imbalance in the reconstructed jets and jet momenta, respectively. On the other hand, also effects caused by the used jet reconstruction algorithm and its efficiency in reconstructing jets have an influence on the momentum imbalance. For example, by using a smaller cone radius less particles of the showers are reconstructed within the jets and therefore the momentum imbalance is enhanced.

In this work, parton showers are not only evolved within vacuum but also within a partonic medium. Therefore a simultaneous treatment of virtual splittings as well as medium-induced scattering processes is in principle necessary. However, because \BAMPS{} is an on-shell transport model, the implementation of virtual splitting processes is not trivial. For that reason we assume for now a separation between the initial virtual splittings within the vacuum and the subsequent in-medium scatterings. For modeling the vacuum splitting we employ partonic events from the event generator \PYTHIA{} (version 6.4.25) \cite{Sjostrand:2006za}. These events are standard \PYTHIA{} events based on leading order pQCD cross sections, where any initial state radiation and primordial $k_t$ is switched off. This allows later a clear comparison between the reconstructed jets and the initial hard scattered partons. For providing a sufficient statistics at high transverse momentum we set the minimum transverse momentum of the initial 
hard \twotwo{} process to $\hat{p}_t>\SI{100}{\GeV}$. The momenta of the ingoing partons are distributed according to CTEQ6L \cite{Pumplin:2002vw} parton distribution functions. The vacuum evolution of the shower partons resulting from the outgoing virtual partons is stopped at the typical hadronic mass scale $Q_0=\SI{1}{GeV}$, which is the default value within \PYTHIA{}. Any subsequent hadronization within \PYTHIA{} is switched off in order to calculate the subsequent in-medium evolution of the parton shower within \BAMPS{}. 

Although the underlying evolution within \BAMPS{} is unchanged, the consideration of vacuum branching processes is in principle conceptually different from our setup used in ref.~\cite{Uphoff:2014cba} and shown in fig.~\ref{fig:raa}, where we used only the unshowered hard scattered partons as initial state for our $R_{AA}$ studies. Neglecting the initial state shower leads to different slopes both in the partonic spectra and thereby also the partonic $R_{AA}$. However, studies show that the subsequent fragmentation into hadrons effectively moderates any potential differences in the resulting $R_{AA}$ of charged hadrons.

As an additional remark, by comparing the $A_J$ distribution of full hadronic \PYTHIA{} \ProtonProton{} events with partonic \PYTHIA{} events one can show that the jet finding procedure effectively removes any influence of the hadronization on the reconstructed leading jet momenta. This is an intended feature of modern jet finding algorithms mainly facilitated by collinear and infrared safety. For that reason we neglect any effects of hadronization on the reconstructed jets throughout this paper. 

\subsection{In-medium evolution of parton showers}\label{sec:evoShower}
In this section we introduce our method for simulating the in-medium momentum loss of reconstructed jets within a partonic transport model. One of the advantages of \BAMPS{} is that the jet energy loss as well as the medium bulk evolution of a heavy-ion collision can be simulated \emph{microscopically} within one common framework. This is different to e.g. energy loss calculations that determine the energy loss by a Monte-Carlo procedure embedded into a macroscopically evolving hydrodynamical background. 

The strategy for simulating the energy loss of parton showers within \BAMPS{} can be separated into the two following parts:
\begin{itemize}
 \item The parton shower events are created by the introduced \PYTHIA{} events on the partonic level. Since these partons are physical particles they have $N_{\text{test}}=1$. This is important to note for the subsequent reconstruction of jets, which can only be done with physical particles. The shower partons are embedded within offline recorded \BAMPS{} events (s. next paragraph) at the spatial insertion point of the corresponding shower-initiating parton pair that is sampled by a Glauber modeling together with a Woods-Saxon density profile. After the standard formation time $\tau_f = \frac{\cosh\left(y\right)}{p_t}$ of the initial partons within \BAMPS{} \cite{Xu:2004mz}, the shower particles are evolved within the recorded \BAMPS{} background events: In every time step the shower partons are allowed to scatter with recorded medium particles via \twotwo{} and \twothree{} processes with the usual stochastic method employed within the \BAMPS{} framework. Whether further scatterings of the recoiled medium 
particles are considered is controlled by two different scenarios that are described in the following sections.
 \item For creating the underlying background event, we simulate heavy-ion collisions with impact parameter $b$ within \BAMPS{} considering full 3+1D expansion and test-particle number $N_{\text{test}}$ with no trigger for high $p_t$ particles. The initial parton distributions are again obtained from \PYTHIA{} together with a Glauber modeling. As it is shown in ref.~\cite{Uphoff:2014cba}, these events show a collective medium behavior quantified in terms of a significant elliptic flow coefficient $v_2$. At every time step we keep track of every scattering process and the phase space information of every particle. With this information it is afterwards possible to ``offline reconstruct'' this event and thereby the evolution of the expanding medium. Similar to most energy loss calculations with hydrodynamical backgrounds, the medium response, which is the modification of the original evolution of the recorded bulk medium by the shower-medium interactions, is therefore not considered throughout this paper. 
Furthermore, the scattering method within \BAMPS{} employing stochastical probabilities is only meaningful while using the introduced test-particle approach. Consequently, scattering processes between shower particles, which are physical particles and therefore have $N_{\text{test}}=1$, are forbidden throughout this paper. In doing so one neglects possible effects of collective behavior between shower particles like e.g. Mach cone structures \cite{Bouras:2014rea}. 
\end{itemize}

\subsubsection{Without recoiled medium partons}\label{sec:norecoil}
For studying the properties of the hot and dense matter created in heavy-ion collisions one is interested in the in-medium momentum loss of high $p_t$ partons created in hard scatterings of nucleons. However, these initial partons decrease their virtuality by vacuum QCD splittings as described in sec.~\ref{sec:vacuum}. Therefore a clear definition of the initial partons is difficult already in \ProtonProton{} collisions and can only be achieved at the moment by reconstructing jets. For calculating the momentum loss of jets after traversing a medium, one has to consider additionally the \twotwo{} scatterings of the initial shower partons with the medium and the gluons emitted via \twothree{} bremsstrahlung processes. This is the procedure applied in most of the current theoretical approaches. Any contribution stemming from recoiled medium partons that subsequently also evolve within the background medium is supposed to be lost within the medium since mostly only macroscopic information about the medium are 
available, which makes the definition of the medium recoil difficult. However, it is possible that this neglected recoil can evolve within in the medium, end up in the reconstructed jets and therefore restore seemingly lost momentum to the reconstructed jets.

However, the procedure of not considering the medium recoil can anyhow only be applied in theoretical approaches, which track the shower particles at every time step and allow thereby a discrimination between shower and medium. In experiments, where this discrimination certainly is not applicable, this procedure is not longer justified. While any contribution to the jets originating from the unscattered background medium is removed by background subtraction, the possible effect of scattered medium particles to the reconstructed jets can survive the subtraction. Consequently, the consideration of recoiled medium partons while comparing theoretical calculations and background subtracted data may become essential.

\subsubsection{With recoiled medium partons}\label{sec:wrecoil}
Since the interactions of shower particles and medium particles are equally treated within \BAMPS{} it is possible to not only investigate the interactions of the initial parton shower but also the role of further interactions of medium particles recoiled in shower-medium interactions. For considering these further in-medium scattering processes, the recoiled medium particles become shower partons themselves. Consequently, the scattered medium particles are allowed to scatter again with other medium particles and evolve like the initial shower partons.

However, these further scatterings can lead to a double counting of scattered medium particles: If the same medium particle is hit twice by the shower particles it would end up more than once as a shower parton. This is an effect caused by a finite number of test-particles: In the limit of infinite test-particle numbers this effect would be naturally cured since the probability of a repeated scattering with the same particle would vanish. To avoid this issue, while employing a finite number of test-particles, we assure that scattered medium particles become only shower partons when they scatter for the first time with a shower particle. Nevertheless, the actual scattering process takes place anyhow and changes the momenta of the outgoing shower parton. This ensures that every scattered medium particle has only one trajectory within the medium and at the same time the effect of the medium on the momenta of the shower partons is preserved. 

When studying the effect of further interactions of the recoiled medium partons on the reconstructed jets, these medium partons can end up in the reconstructed jets and therefore transport medium momentum to the jets. Therefore it is essential to consider an appropriate subtraction of the background momentum from the jets before comparing with experimental data. The experimental subtraction algorithms consist mostly of an estimation of the average background momentum density within the size of a jet cone that is finally subtracted from the reconstructed jet momenta. However, when reconstructing jets based on only the parton shower together with the recoiled medium particles these jets possess only a part of the background momentum. The missing, even larger part of medium momentum that should end up within the reconstructed jets comes from medium particles that did not scatter with the shower. Therefore, due to our chosen separation of shower and background event, we first have to appropriately combine the 
shower and the not scattered medium particles of our simulation to form full ``shower+medium'' events before employing experimental subtraction procedures. After the combination we are able to estimate correctly the underlying background momentum. 

However, within \BAMPS{} the test-particle scaling of particle numbers complicates the situation: Jet reconstruction is obviously only reasonable if it is done event-by-event based on physical particles. This requirement is not fulfilled for the medium particles that were not scattered by the shower particles. Due to the employed test-particle ansatz, the medium particles are test-particles ($N_{\text{test;medium}} \neq 1$) while the shower particles are physical particles ($N_{\text{test}}=1$). Thus, after simulating the shower and before combining the two regimes we first have to rescale both test-particle numbers to the same value. After this rescaling, only {\em physical} medium particles are left within the background event, which enables us to combine those with the shower partons to form full $N_{\text{test}}=1$ ``shower+medium'' events. For more details about the scaling of test-particles and thereby generating appropriate background events see appendix~\ref{app:sampling}, where also the validity of 
this scaling is shown.

Neglecting any hadronization effects, the so obtained events are similar to experimentally measured events. To subtract finally the background contamination from reconstructed jets based on parton showers including recoiled medium partons, we employ the ``CMS noise/pedestal subtraction method'' \cite{Kodolova:2007hd}. This algorithm is an iterative procedure that first estimates the background transverse momentum, then reconstructs jets, excludes the reconstructed jets from the background estimation and again estimates the background. After some iterations the so obtained average background transverse momentum is subtracted from the finally reconstructed jets. For more information about this subtraction algorithm we refer to ref. \cite{Kodolova:2007hd}.

\subsection{Consideration of detector effects}
\label{sec:smearing}
Naive comparisons of the $A_J$ distribution of hadronic \PYTHIA{} events with the $\sqrt{s}=\SI{2.76}{\tera\electronvolt}$ \ProtonProton{} $A_J$ distribution measured by CMS \cite{Chatrchyan:2011sx} show that \PYTHIA{} cannot describe the measured data out of the box. This is caused by the effect that the detectors themselves have only a finite resolution that leads to fluctuations in the reconstructed jets and by that to an enhancement of momentum imbalance absent in the simulations.

As a first step for modeling the detector effects we reconstruct jets within this paper at ``calorimeter level'' and not at ``particle level''. This is done by employing a ``calorimeter'', which is modeled by a grid in rapidity $y$ and azimuthal angle $\phi$, in which the final particles of each simulation are sorted depending on their $y$ and $\phi$. The size of the grid cells is based on the cell size of the CMS calorimeters. Moreover, we employ no trigger condition for the individual particle transverse momentum like it should be done in principle when comparing with CMS data because of their strong magnetic field prohibiting particles with $p_{t}\lesssim\SI{1}{\GeV}$ to reach the calorimeters. However, due to the missing hadronization a comparison between hadronic momentum cuts of CMS and partonic momentum cuts necessary within BAMPS are difficult.

Moreover, motivated by theoretical calculations \cite{Qin:2010mn} and the detector response analysis by CMS \cite{CMS:2011ab}, we use the following strategy for mimicking additional detector effects: We apply an independent smearing procedure on the leading and subleading jet momenta $p_{t;1}$ and $p_{t;2}$, which is based on a Gaussian function $N\left(p_{t;i}\,,\,c\sqrt{p_{t;i}}\right)$ with width $\sigma=c\sqrt{p_{t;i}}$ and fit parameter $c$. This smearing procedure alters the two jet momenta independently, which leads to an additional amount of momentum asymmetry within one event. For smooth $A_J$ distributions before smearing, the statistics of the simulations is enhanced by repeating the stochastic smearing $N_{\text{smeared}}=\num{1000}$ times for each event. 

\begin{figure}[tb]
 \includegraphics[width=\linewidth]{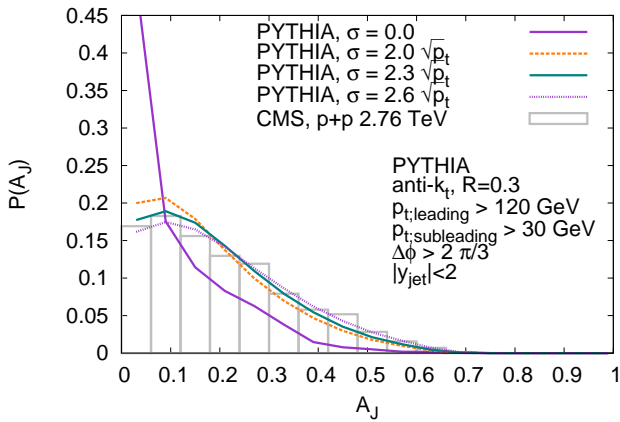}
 \caption{(Color online) $A_J$ distribution for \PYTHIA{} events with different smearing parameters in comparison with $\sqrt{s}$ = \SI{2.76}{\tera\electronvolt} \ProtonProton{} data \cite{Chatrchyan:2012nia}. All experimental trigger conditions as defined in ref.~\cite{Chatrchyan:2012nia} and mentioned in Sec.~\ref{sec:results} are applied.} 
 \label{fig:Ajpp}
\end{figure}

The value of the smearing parameter $c$ is determined by fitting the $A_J$ distribution of \PYTHIA{} simulations on the partonic level to the $\sqrt{s} = \SI{2.76}{\tera\electronvolt}$ CMS \ProtonProton{} data \cite{Chatrchyan:2012nia} as shown in fig.~\ref{fig:Ajpp}. The agreement between the two distributions is best for a smearing factor $c = \num{2.3}$. The Gaussian smearing with the so determined smearing factor $c$ is used in the following study as an effective detector filter for the \PbPb{} results shown in Sec.~\ref{sec:results}. As a remark, although the CMS collaboration provides a parametrization for the jet energy resolution \cite{Chatrchyan:2012gt}, this is not applicable here since we are dealing with \PYTHIA{} events without considering any hadronization and resonance decay effects. 

\section{Simulation results}
\label{sec:results}
After introducing our simulation strategy for parton showers within \BAMPS{} in the previous sections, we present our results for the momentum loss of reconstructed jets in terms of the momentum imbalance $A_J$. For this study we use full 3+1D, expanding \BAMPS{} heavy-ion simulations of $\sqrt{s}=\SI{2.76}{\ATeV}$ \PbPb{} collisions with impact parameter $b_{\text{mean}} = \SI{3.4}{\fm}$ that corresponds to an experimental centrality class of \SIrange{0}{10}{\percent} as determined in Monte Carlo Glauber calculations \cite{Chatrchyan:2011sx}. The embedded parton showers are initialized by \PYTHIA{} initial conditions and afterwards evolved within the medium as described in Sec.~\ref{sec:simStrategy}. Possible detector effects are considered by the methods described in Sec.~\ref{sec:smearing}. For the comparison with experimental data we employ for all results in this section the trigger conditions defined by CMS for their $A_J$ studies. We use the anti-$k_t$ algorithm with a distance parameter $R=0.3$ as 
provided by the FastJet package \cite{Cacciari:2011ma}. For fulfilling the trigger conditions by CMS \cite{Chatrchyan:2012nia}, the events used in the following have a leading jet with $p_{t;leading}>\SI{120}{\giga\electronvolt}$ and a subleading jet with $p_{t;subleading}>\SI{30}{\giga\electronvolt}$. Both jets have to be close to mid-rapidity ($|y_{\text{Trigger;Jets}}| < \num{2.0}$) and the difference between their azimuthal angle is limited to $\Delta \phi_{\text{Trigger;Jets}} > 2\pi/3$.

\subsection{Momentum asymmetry of reconstructed jets within BAMPS}
\begin{figure}[tb]
 \includegraphics[width=\linewidth]{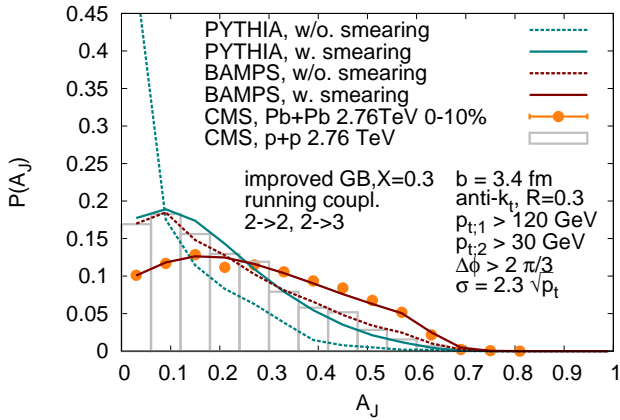}
 \caption{(Color online) $A_J$ distribution of \BAMPS{} events with $b_{\text{mean}}=\SI{3.4}{\fm}$ based on \PYTHIA{} initial conditions together with the initial \PYTHIA{} distribution in comparison with $\sqrt{s}=\SI{2.76}{\tera\electronvolt}$ \ProtonProton{} and $\sqrt{s}=\SI{2.76}{\ATeV}$ 0-10\% \PbPb{} data measured by CMS \cite{Chatrchyan:2012nia}. With dashed lines the respective $A_J$ distributions before considering detector effects by a Gaussian smearing are shown.}\label{fig:Aj}
\end{figure}

Fig.~\ref{fig:Aj} shows the momentum imbalance $A_J$ as calculated within \BAMPS{} while neglecting any effect caused by recoiled medium partons. For better understanding the in-medium momentum loss, the initial momentum imbalance distribution before evolving within \BAMPS{} and as obtained by \PYTHIA{} is also shown. The dashed lines depict the respective $A_J$ distributions before smearing the jet momenta for modeling detector effects, while the solid lines show the distributions after smearing. Due to the independent smearing of the leading jet and the subleading jet, the smearing procedure enhances the momentum imbalance significantly both in vacuum and in heavy ion collisions. As shown in fig.~\ref{fig:raa} in sec.~\ref{sec:bamps} the suppression of single inclusive hadrons in terms of the nuclear modification factor $R_{AA}$ at LHC is described by \BAMPS{} calculations when using a combination of the improved GB matrix element, the running coupling evaluated on a microscopic level, and an effective LPM 
implementation with 
parameter $X_{\text LPM}=0.3$, which was chosen in a comparison to RHIC $R_{AA}$ data \cite{Uphoff:2014cba}. While employing the same setup also for the studies concerning the momentum imbalance $A_J$ of reconstructed jets, \BAMPS{} calculations show an excellent agreement with data.

\begin{figure}[htb]
 \subfloat[][]{\includegraphics[width=\linewidth]{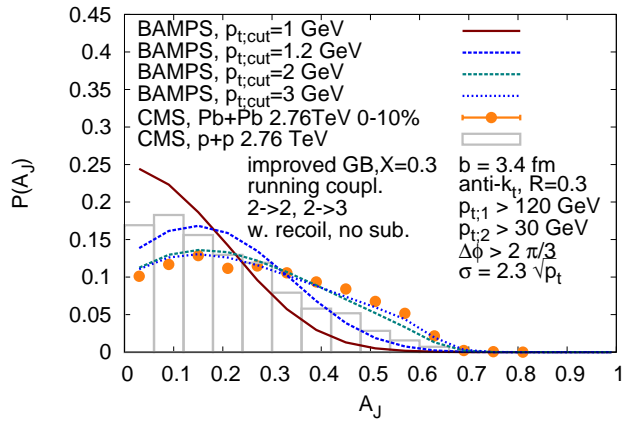}\label{fig:Aj_recoil_noSub}}

 \subfloat[][]{\includegraphics[width=\linewidth]{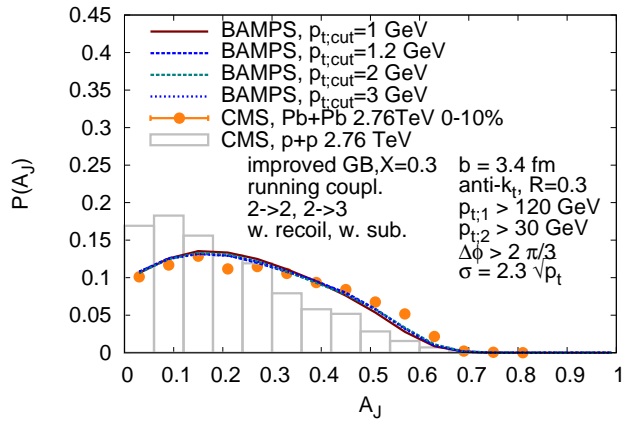}\label{fig:Aj_recoil_wSub}}
 \caption{(Color online) $A_J$ distribution of \BAMPS{} events with $b_{\text{mean}}=\SI{3.4}{\fm}$ based on \PYTHIA{} initial conditions in comparison with $\sqrt{s}=\SI{2.76}{\tera\electronvolt}$ \ProtonProton{} and $\sqrt{s}=\SI{2.76}{\ATeV}$ 0-10\% \PbPb{} data measured by CMS \cite{Chatrchyan:2012nia}. Shown are distributions of $A_J$ for reconstructed jets while considering also the fully evolved recoiled medium particles without considering any background subtraction (fig.~\ref{fig:Aj_recoil_noSub}) and with considering background subtraction (fig.~\ref{fig:Aj_recoil_wSub}) for different transverse momentum cuts  $p_{t;cut}$ in the shower evolution.}\label{fig:Aj_recoil}
\end{figure}

In order to investigate the role of recoiled medium partons for the momentum imbalance of reconstructed jets, we show in fig.~\ref{fig:Aj_recoil} $A_J$ distributions for simulations in which the further evolution of recoiled medium partons is now explicitly considered. As a measure for differentially investigating the effect of the further in-medium evolution of the recoiled medium particles we terminate the evolution of the shower and recoiled medium partons when their transverse momentum reaches a minimum value $p_{t;cut}$. Partons with less transverse momentum stream freely without any scattering process until the simulation ends. This cut in the parton evolution additionally minimizes the numerical effort necessary for calculating the evolution of many very low $p_t$ particles that are supposed to be subtracted in the experimental analyses anyhow since they are part of the background.

As shown in fig.~\ref{fig:Aj_recoil_noSub}, where no background subtraction is applied, the influence of recoiled medium partons is minor if their evolution is stopped at a higher transverse momentum cuts $p_{t;cut}>\SI{2}{\GeV}$. However, when the partons are allowed to evolve to lower $p_t$, the $A_J$ distribution are more and more shifted to lower momentum imbalances. This effect stems from low momentum, recoiled medium partons that even after further evolving within the medium stay within the jet cones and do not contribute to a momentum loss out of the cones as it is observed in the simulation without recoiled medium particles. In other words: If the parton shower loses momentum by shower-medium interactions, a part of the momentum stays in the reconstructed jets as recoiled medium partons, contaminates the jets with background momenta, and thereby decreases the resulting observed jet momentum loss. Therefore, in order to estimate the genuine contribution of the recoiled medium to the jets, a background 
subtraction as described in sec.~\ref{sec:wrecoil} and used in experimental analyses becomes essential. As shown in fig.~\ref{fig:Aj_recoil_wSub}, this subtraction effectively moderates the effect of recoiled low $p_t$ medium partons on the momentum imbalance $A_J$ and thereby leads to coinciding $A_J$ distributions for the different $p_{t;cut}$ values. Furthermore, the momentum imbalance of parton showers with recoil and background subtraction agrees with the result for parton showers without recoil and background subtraction. This finding supports on one hand the applicability of experimental background subtraction methods by fully microscopic simulations. On the other hand it also justifies the procedure already applied in other theoretical models that neglect the contribution of further in-medium scatterings of the recoiled medium particles (see fig.~\ref{fig:Aj}). Since neglecting the recoiled medium particles simplifies the simulations tremendously we present therefore in the following only results 
without recoiled medium partons and without subsequently subtracting the background.

\subsection{Relation between initial partons and reconstructed jets}
Simulations within transport models benefit from the advantage of providing the full microscopic phase space information of every particle underlying the jet reconstruction. Such studies lead to a further understanding of the mechanisms causing the momentum loss. For that reason we investigate the momentum loss of reconstructed jets by comparing the final jet transverse momenta after traversing the medium with the transverse momenta of the initial hard scattered partons from \PYTHIA{}. To this end, we define the momentum loss $\Delta p_t = p_{\text{t;jet}} - p_{\text{t;init. parton}}$ depending on the initial parton transverse momentum for the leading and subleading jets, respectively. As a note, since we are interested in the genuine momentum loss of the jets, we investigate within this section events in which the respective smeared jets fulfill the experimental trigger conditions by CMS but show the respective momentum loss based on these jets before smearing.

\begin{figure}[tb]
 \includegraphics[width=\linewidth]{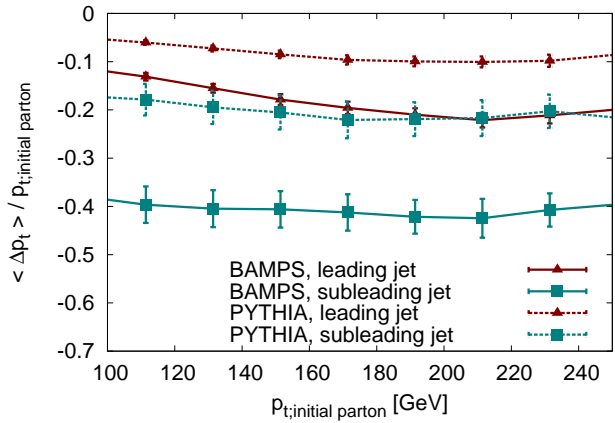}
 \caption{(Color online) Average momentum loss of reconstructed jets compared to their initial shower-initiating partons $\Delta p_t = p_{\text{t;jet}} - p_{\text{t;init. parton}}$ in relation to the initial parton transverse momentum. Dashed lines show the momentum loss already present due to vacuum splittings by \PYTHIA{}, while the solid lines show the momentum loss after evolution within \PYTHIA{} and \BAMPS{}. For the underlying events all experimental trigger conditions as described in Sec.~\ref{sec:results} are applied.}
\label{fig:ptDPt}
\end{figure}

Fig.~\ref{fig:ptDPt} shows the average momentum loss $\langle\Delta p_t\rangle$ as a function of the initial parton $p_{\text{t;init. parton}}$. While the dashed lines depict the momentum loss within \PYTHIA{} due to the initial vacuum splittings, the solid lines show the overall momentum loss due to both the initial splittings and the subsequent in-medium evolution. The relative momentum loss of both the leading and subleading jet is approximately constant over the whole considered $p_t$-range: While the subleading jet loses on average approx. \SI{40}{\percent} of its momentum, the leading jet loses approx. \SI{20}{\percent} of its momentum with respect to its initial shower-initiating parton. Thereby both jets lose on average already \SI{10}{\percent} (leading jet) and \SI{20}{\percent} (subleading jet) of its initial parton momentum due to vacuum splittings.

\begin{figure}[tb]
 \includegraphics[width=\linewidth]{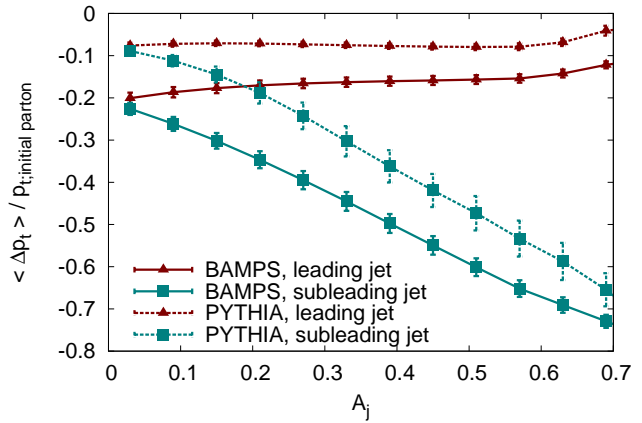}
 \caption{(Color online) Average momentum loss of reconstructed jets compared to the initial shower-initiating parton $\Delta p_t = p_{\text{t;jet}} - p_{\text{t;init. parton}}$ in relation to the momentum asymmetry $A_J$ of this respective event. Again, dashed lines show the momentum loss already present due to vacuum splittings by \PYTHIA{}, while the solid lines show the momentum loss after evolution within \PYTHIA{} and \BAMPS{}. For the underlying events all experimental trigger conditions as described in Sec.~\ref{sec:results} are applied.}
\label{fig:AjDpt}
\end{figure}

Furthermore, the momentum loss of reconstructed jets with respect to their initial partons is studied in bins of the momentum asymmetry $A_J$ as shown in fig.~\ref{fig:AjDpt}. Again the solid lines show the momentum loss after the in-medium evolution and initial vacuum splittings, while the dashed lines depict the momentum loss due to only \PYTHIA{}. The momentum loss of the subleading jet is for only vacuum splittings as well as after traversing the medium a decreasing, almost linear function of $A_J$. In contrast, the momentum losses of the leading jet after the vacuum and after traversing additionally the medium is almost independent from the underlying momentum imbalance. This means that the observed momentum imbalance within \BAMPS{} is mainly caused by the momentum loss of the subleading jet. However, the relative momentum loss of both jets by the in-medium evolution is approximately the same over the whole $A_J$ range: While at small momentum imbalances both the leading and subleading jet momentum 
loss drops from approx. \SI{10}{\percent} to approx. \SI{20}{\percent} after traversing the medium, at higher imbalance values $A_J\approx0.6$ the relative drop of both jets remains unaltered at approx. \SI{10}{\percent} but the overall momentum loss of the subleading jet is with \SI{70}{\percent} much higher.

Besides the momentum information of the initial particles, also their spatial positions are available in \BAMPS{} simulations through the whole evolution. For this reason it is possible to investigate microscopically the underlying differences in path length dependence responsible for the respective momentum imbalance. Together with the creation vertices of the initial partons and their initial direction in the transverse plane, one can geometrically define a distance between the creation point of the parton pair and the point where the partons leave the initial, almond-shaped collision zone. For more details about this definition of the transverse in-medium path length cf. Appendix~\ref{app:length}. By comparing the two distances $L_{l/s}$, where $L_l$ ($L_s$) is the longer (shorter) transverse in-medium path length of the two partons, one can define analogously to the momentum imbalance $A_J$, the transverse length imbalance
\begin{align}
 L_i = \frac{L_l - L_s}{L_l + L_s} \, .
 \label{eq:Li}
\end{align}
Lower $L_i$ values correspond to events in which the two initial partons have to traverse a similar transverse distance within the medium, while events with higher $L_i$ are events in which one of the partons traverses more medium than the other in the transverse direction. 

\begin{figure}[tb]
 \includegraphics[width=\linewidth]{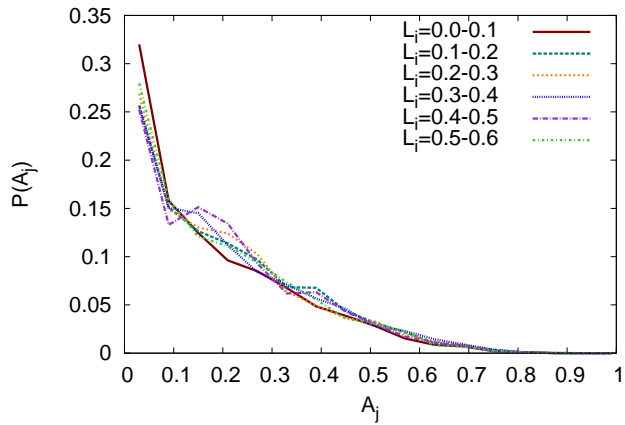}
 \caption{(Color online) Momentum imbalance $A_J$ in bins of the length imbalance $L_i$ (Eq.~\ref{eq:Li}). For the underlying events all experimental trigger conditions as described in Sec.~\ref{sec:results} are applied.}
 \label{fig:AjLi}
\end{figure}

\begin{figure}[tb]
 \includegraphics[width=\linewidth]{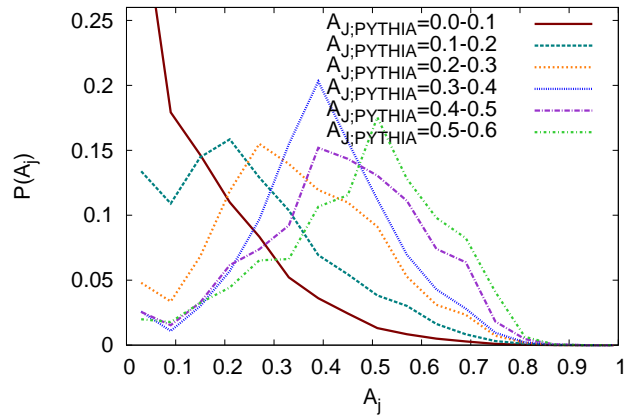}
 \caption{(Color online) Momentum imbalance $A_J$ in bins of the momentum imbalance $A_{J;\text{\PYTHIA{}}}$ already present before the in-medium evolution. For the underlying events all experimental trigger conditions as described in Sec.~\ref{sec:results} are applied.}
 \label{fig:AjAj}
\end{figure}

Fig.~\ref{fig:AjLi} shows the momentum imbalance for different length imbalance ranges based on calculations without considering the medium recoil and background subtraction. As a note, again the genuine distributions before smearing are shown based on events in which the smeared jets pass the trigger conditions. The different $A_J$ distributions show almost no dependence on the underlying path-length difference! Therefore the tomographic capability of $A_J$ seems to be limited. A similar observation has also been done in ref.~\cite{Renk:2012cx}. Much more relevant for the resulting momentum imbalance is the momentum imbalance of the respective event that is present before evolving within the medium. This can be seen in fig.~\ref{fig:AjAj}, where again different $A_J$ distributions are shown but this time binned in the event-by-event momentum imbalance $A_{J;\text{PYTHIA}}$ already present in the vacuum before evolving in the \BAMPS{} medium. In contrast to the binning in $L_i$ these distributions show a 
clear dependence of the ``heavy-ion $A_J$'' from the ``vacuum $A_{J;\text{PYTHIA}}$''.

The non-sensitivity of the resulting momentum imbalance on the underlying path length difference can be understood by two different, generic and model independent considerations. On one hand it can happen that the initial leading and subleading jet switch roles while traversing the medium: the jet that was the leading jet before the medium evolution becomes the subleading jet after traversing it and vice-versa. This makes the resulting momentum imbalance rather insensitive to the underlying path length dependence. As an example, let us assume that the initial leading jet has $p_{t;leading}^i=\SI{130}{\GeV}$ and loses \SI{60}{\GeV} while traveling a large distance within the medium with the result that it ends up as the subleading jet with $p_{t;subleading}^f=\SI{70}{\GeV}$. The initial subleading jet starts with a momentum of $p_{t;subleading}^i=\SI{100}{\GeV}$ (it has lost $\geq\SI{30}{\GeV}$ by vacuum splittings), travels a very small distance within the medium and loses thereby no momentum and 
therefore becomes the final leading jet $p_{t;leading}^i=\SI{100}{\GeV}$. But this means that the difference between the initial $A_J^i=0.13$ and final $A_J^f=0.17$ momentum imbalances almost vanishes while the difference in the in-medium path lengths is significant. On the other hand, even if the jets do not switch roles the momentum imbalance loses its tomographic capability when an initial momentum asymmetry is present as it was also observed in ref.~\cite{Uphoff:2013rka}, where heavy quark correlations were studied. Assuming an initial momentum asymmetry between the initial leading and subleading jet momenta, the momentum imbalance decreases when the leading jet loses momentum. However, when the subleading jet loses momentum, the momentum imbalance increases. Both effects counteract each other with the result that the momentum imbalance seems to be independent from the underlying difference in the in-medium momentum losses but are much more related to the initial, vacuum momentum imbalance.

\section{Summary}
Within this paper we presented results on reconstructed jets within the partonic transport model \BAMPS{}. Heavy-ion simulations by \BAMPS{} benefit from the possibility to calculate the full evolution of both the high $p_t$ particles and the softer bulk medium particles in one common approach while employing the same microscopic interactions. This allows not only the investigation of the actual parton shower traversing the heavy-ion medium but also the evolution of the recoiled medium particles within the same approach. 

While employing an improved version of the Gunion-Bertsch matrix element, a microscopically evaluated running coupling and an effective LPM implementation, we show that the momentum loss of high $p_t$ parton showers in terms of the momentum imbalance $A_J$ is in agreement with data when not considering recoiled medium partons. This supports former studies, in which the nuclear modification factor $R_{AA}$ was investigated within the same setup and agreed well with experimental observations.

Furthermore, we then investigated in detail the influence of the further in-medium evolution of recoiled medium partons on the momentum imbalance $A_J$. We found that the influence of the recoiled medium is limited to momentum regions $p_t<\SI{2}{\GeV}$ and only if no additional background subtraction is applied. This means that the background subtraction effectively eliminates the influence of recoiled medium partons on the momentum imbalance what in addition implies that it is sufficient to study the momentum imbalance $A_J$ by only considering the momentum loss of the initial parton shower and subsequent bremsstrahlung gluons. Furthermore, this agreement additionally supports the applicability of background subtraction methods as used by experiments from a theoretical point of view.

Besides the momentum imbalance $A_J$ we studied the relation between the reconstructed jets and their initial shower-initiating partons while employing all experimental trigger conditions. Independent from the initial parton momentum, the reconstructed subleading jet loses $\approx \SI{40}{\percent}$ of momentum with respect to its initial parton momentum due to initial vacuum splittings and the subsequent in-medium evolution. Investigations of the momentum loss causing the momentum imbalance showed that the leading and subleading jet loses an equal amount of momentum within the medium, while the momentum imbalance is nearly exclusively caused by the subleading momentum loss. Also the sensitivity of the momentum imbalance $A_J$ on the underlying difference in the in-medium path length seems to be limited. On the contrary, the final momentum imbalance after traversing the medium is dominated by the initial momentum imbalance after the vacuum evolution.

We plan for the future, to study within our approach also other jet-related observables like e.g. the suppression of reconstructed jet spectra. In addition, we will revisit our current implementation of the LPM effect and study the potential applicability of its stochastic implementation \cite{Zapp:2011ya,Zapp:2012ak}.

\begin{acknowledgments}
F.S. wants to thank Bj\"orn Schenke for stimulating and helpful discussions regarding the combination of shower and background events. Furthermore, we want to thank the Helmholtz Graduate School for Heavy-Ion Research (HGS-HIRe) and the Helmholtz Research School for Quark Matter Studies (H-QM) for financial support. This work was supported by the Bundesministerium f\"ur Bildung und Forschung (BMBF) and by the Helmholtz International Center for FAIR within the framework of the LOEWE program launched by the State of Hesse. Z.X. is supported by the NSFC under grant No. 11275103. The \BAMPS{} simulations were performed at the Center of Scientific Computing (CSC) of the Goethe-University Frankfurt. 
\end{acknowledgments}

\appendix
\section{Discussion of test-particle corrected background events}
\label{app:sampling}
\subsection{Sampling of test-particle corrected background events}
As described in Sec.~\ref{sec:simStrategy}, due to the employed test-particle ansatz one has to rescale appropriately \BAMPS{} background events ($N_{\text{test}}\neq1$) before combining them with shower events ($N_{\text{test}}=1$). Therefore it is necessary to determine which medium particle is physical and which is a test-particle. This discrimination is only achievable by stochastical methods: Whether a particle is physical or a test-particle is decided by Monte-Carlo sampling. 

Before going to the actual scenario of events with shower and medium interactions, we first discuss the sampling probabilities of a simpler scenario of \BAMPS{} events without any additional shower partons. These events only consist of medium particles with test-particle number $N_{\text{test}}\neq 1$ and a combination of shower and medium part, which complicates the sampling procedure, is not needed. Within this simple scenario, the probability for a medium particle to be physical is equally distributed among the medium particles. Therefore, this stochastical probability obviously reads
\begin{align}
 P_{\text{ref}} = \frac{N_{\text{physical}}}{N} = \frac{1}{N_{\text{test}}} \, ,
\end{align}
where $N_{\text{physical}}$ is the number of physical particles, $N$ the total number of particles and $N_{\text{test}}$ as before the number of test-particles per physical particles in this event. We will denote this simple sampling case in the following as the {\em ``reference sampling method''}.

In contrast, when considering events in which the medium particles are scattered by shower partons, the described reference sampling scenario and its probabilities cannot be applied anymore: Some of the medium particles have interacted with the shower during the simulation and became therefore ``shower partons'' themselves. If one simply uses the reference sampling probabilities, some scattered medium particles could be sampled successfully as physical medium particles and would end up therefore as a sampled medium particle as well as a shower particle in the same ``shower+medium'' event. To avoid this possible double counting, only medium particles that were not scattered by any shower parton are considered in the sampling process and may become physical medium particles. 

However, the {\em exclusion} of scattered medium particles from the sampling procedure leads to an overestimation in the sampling probabilities of the unscattered particles: Because there are already $N_{\text{scatt}}$ scattered, physical medium particles, which were ``chosen'' during the shower simulation, only $N_{\text{physical}}-N_{\text{scatt}}$ particles are still allowed to become physical. Thus this exclusion has to change the sampling probabilities and becomes especially important in regions in which many scattered medium particles are present. To avoid the overestimation of medium particles, the sampling probabilities in events with added shower partons have to be modified depending on the number of scattered---and thereby already physical---medium particles nearby in phase space. To account for this local dependence, the probabilities are calculated based on equally sized cells in rapidity $y$ and azimuthal angle $\phi$. This choice of cells is motivated by thinking of the experimental jet 
reconstruction based on 
calorimeter cells. 
The probability for an unscattered medium particle to be physical within such a cell then reads
\begin{align}
 P_{\text{local}} = \frac{N_{\text{physical;cell}}-N_{\text{scatt;cell}}}{N_{\text{cell}}-N_{\text{scatt;cell}}} = \frac{N_{\text{cell}}/N_{\text{test}}-N_{\text{scatt;cell}}}{N_{\text{cell}}-N_{\text{scatt;cell}}} \, ,
 \label{eq:Plocal}
\end{align}
where $N_{\text{physical;cell}}=N_{\text{cell}}/N_{\text{test}}$ is the number of physical particles, $N_{\text{cell}}$ the total number of medium particles---including scattered and unscattered---and $N_{\text{scatt;cell}}$ the number of scattered medium particles within the respective cell. We will denote this sampling method in the following as the {\em ``local sampling method''}. The size of the used cells is a crucial ingredient for the sampling procedure: As already visible in Eq.~\ref{eq:Plocal} a too small cell size could lead to divergent or negative probabilities, while a too coarse cell size cannot correctly resolve the shower region in the $y$-$\phi$ space.

After sampling for every unscattered medium particle whether it is physical or not, we end up with a pure $N_{\text{test}}=1$ background event without any scattered medium particle. For obtaining the final events we add these sampled medium particles to our previously simulated physical partons, which consist of both the initial shower partons and the recoiled medium partons. In this way we attain finally an event with only physical particles left. Based on these events we are now able to calculate the average transverse momentum density and thereby employ experimental subtraction methods.

\subsection{Comparing different sampling methods}
To check our proposed rescaling techniques we want to compare in this appendix section events based on the local sampling method with events sampled by the reference sampling. The rescaling of events to $N_{\text{test}}=1$ will introduce uncertainties into the analysis of reconstructed jets. Before studying the subtracted jets it is crucial to estimate these additional uncertainties of the rescaling algorithms. 

For investigating these uncertainties we study the summed transverse momentum within cones in $y-\phi$ that consist of only $N_{\text{test}}=1$ medium particles. To create such test events we use offline recorded \BAMPS{} events that were traversed by a parton shower (cf. Sec.~\ref{sec:simStrategy}). Afterwards we sample for each unscattered medium particle whether it is physical or not by the local sampling method. Instead of adding the sampled physical medium particles to the shower partons to obtain a $N_{\text{test}}=1$ event, we subsequently combine in this section the sampled medium particles with the scattered medium particles before their respective interaction with the shower. Thus we should end up with events with only $N_{\text{test}}=1$ medium particles while no original shower partons are present. 

The so obtained events can be compared to the pure background events in which no shower was present and therefore the reference sampling method is applicable. As already pointed out in the previous section, the reference sampling is the correct method for scaling down the number of test-particles in events without any additional shower. Thus if we compare the locally sampled pure medium event to the sampled medium events with the reference method we get an estimate of the effects of the sampling procedure on the amount of transverse background momentum.

Because our main goal is to create an appropriate jet background we investigate the uncertainties in terms of background jet momenta. Since this sampling study should be understood as rather qualitative, it is assumed to be sufficient to calculate the transverse momentum within fixed jet cones instead of fully reconstructing jets. Therefore we neglect additional effects to the background momenta introduced by the full reconstruction algorithms. The energies and momenta of the jet cones are evaluated by summing up all four-momenta of sampled medium particles, which are located inside a cone with radius $\Delta R = \sqrt{\Delta \phi^2 + \Delta y^2}$ in the $y-\phi$ plane around a fixed axis. This choice of simple fixed cone ``jet definition'' is motivated by the first-generation jet cone algorithms, which were based on the idea of a conserved direction of energy flow. 

Since we are interested in the different effects of the sampling procedures on the background cone momenta for the shower and the non-shower region, we investigate two different jet cone axes: an axis directed in the initial hard parton direction denoted as the ``in-shower direction'' and an axis with same azimuthal direction but different rapidity ($y = 3.5$) denoted as the ``out-of-shower direction''. This different rapidity ensures the investigation of a region not traversed by the parton shower. Intuitively, in the direction where the initial parton has flown the density of scattered medium particles should be higher than in the out-of-shower direction. Therefore we expect differences in the cone momenta between the two regions if the proposed sampling algorithms are not sufficiently effective.

\begin{figure}[tb]
        \subfloat[][]{\includegraphics[width=\linewidth]{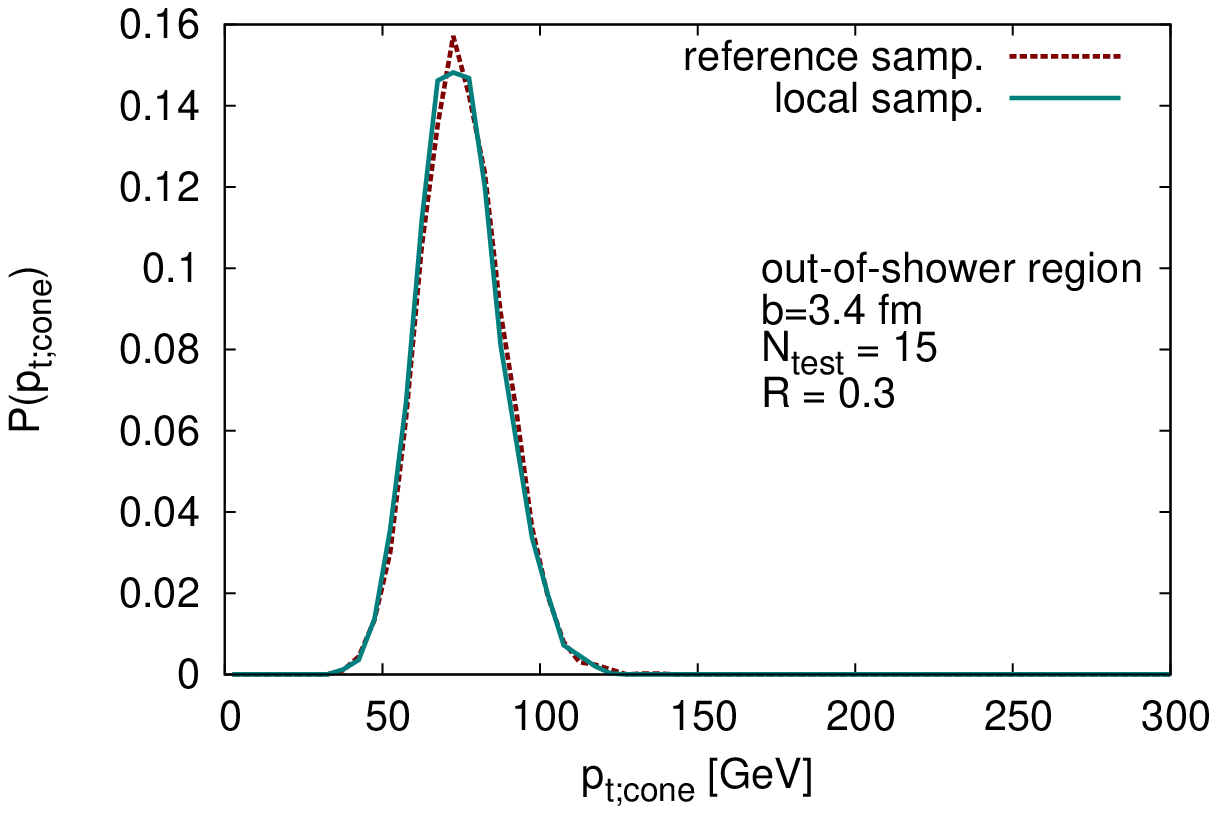}\label{fig:coneOutOfShowerPt}}

	\subfloat[][]{\includegraphics[width=\linewidth]{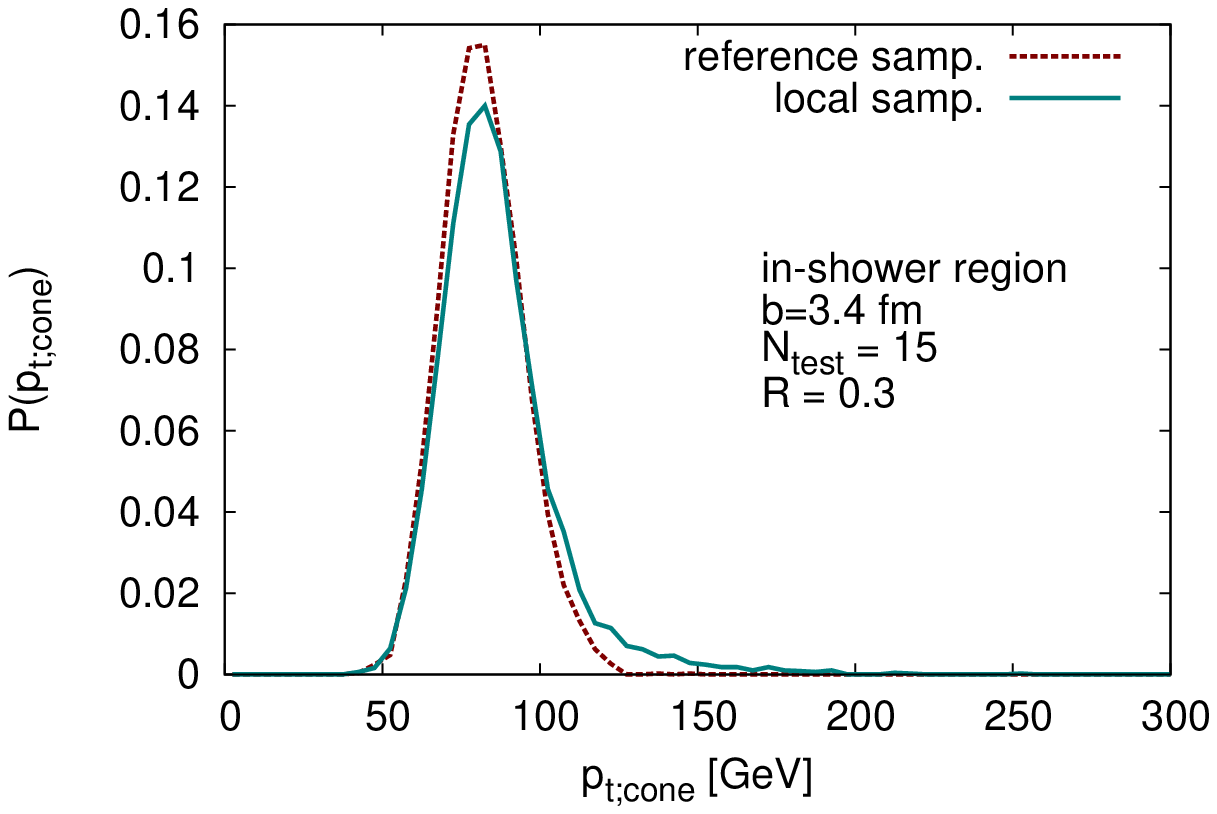}\label{fig:coneInShowerPt}}
        \caption{(Color online) Distribution of the background momentum $p_{t;cone}$ within a cone of $\Delta R=0.3$ for the reference and the local sampling method. Shown are the distributions for the out-of-shower direction (fig.~\ref{fig:coneOutOfShowerPt}) and the in-shower direction (fig.~\ref{fig:coneInShowerPt}).}\label{fig:coneShowerPt}
\end{figure}

Fig.~\ref{fig:coneShowerPt} shows the distribution of summed background transverse momentum within a cone with $\Delta R = 0.3$ placed in either the out-of-shower (fig.~\ref{fig:coneOutOfShowerPt}) or in-shower direction (fig.~\ref{fig:coneInShowerPt}). For the offline recorded, underlying background event we use $\sqrt{s}=\SI{2.76}{\TeV}$ \PbPb{} heavy-ion collisions simulated with \BAMPS{} with $N_{\text{test}}=15$ and impact parameter $b=\SI{3.4}{\fm}$, which corresponds to \SIrange[tophrase=-]{0}{10}{\percent}-centrality at LHC. The reference sampling case shows a Gaussian distribution for the background cone momenta with mean  $p_{\text{t;background}} \approx \SI{75}{\GeV}$ and width $\sigma \approx \SI{25}{\GeV}$ in both cone-axis directions. This independence from the cone axis is, of course, expected because in this reference case no traversing shower is present and thus there is no different in- and out-of-shower direction. 

As seen in fig.~\ref{fig:coneShowerPt} both sampling methods, the ``local sampling'' method and the ``reference sampling'' method, show the same background cone momenta distributions for both the in-shower and out-of-shower region. Due to our introduced modification of the sampling probabilities within the ``local sampling'' method, any possible overweighting of sampling probabilities while considering scattered medium particles is cured. Furthermore, independently from the employed sampling method, background events within \BAMPS{} have a $p_{\text{t;background}}\approx\SI{75}{\GeV}$ within a fixed cone of $R=0.3$. Although this sampling analysis should be understood as rather qualitative because of the used jet definition we are now able to conclude that the introduced uncertainties in the transverse momentum of cones with size $\Delta R=\num{0.3}$ when sampling the corrected background are modest for the in- as well as the out-of-shower region. Under the assumption that the effects following from a proper 
jet reconstruction with, for example, the anti-$k_t$-algorithm should be also modest, our qualitative statement for the sampling of particles will hold.

\section{Definition of in-medium path lengths\label{app:length}}
For studying the dependence of the momentum imbalance $A_J$ on the difference of the in-medium path lengths of the initial partons we defined in Sec.~\ref{sec:results} the transverse length imbalance parameter $L_i$. To this end, a clear definition of the transverse in-medium path lengths is essential. Because \BAMPS{} provides full phase space information including the spatial coordinates of every particle, one could in principle track the correct in-medium path length of each single parton. However, the definition of a path length of an object consisting of multiple particles, like e.g. a reconstructed jet, is difficult. Since the length imbalance $L_i$ measures only the relative difference between the path lengths, it is though possible to define the distances traveled by the partons not in terms of the actual in-medium path lengths but as the distance from the creation point of the initial parton pair to the edge of the collision zone $L_{l/s}$, in direction of the initial transverse parton momentum. 
Neglecting any transverse expansion, the difference between those distances estimates qualitatively the subsequent difference of the in-medium path lengths. Because we are interested in jets traversing the medium at close to mid-rapidity ($|y|<2$), we neglect, as a first estimate, any longitudinal path length dependence and focus on the trajectories in the transversal plane.

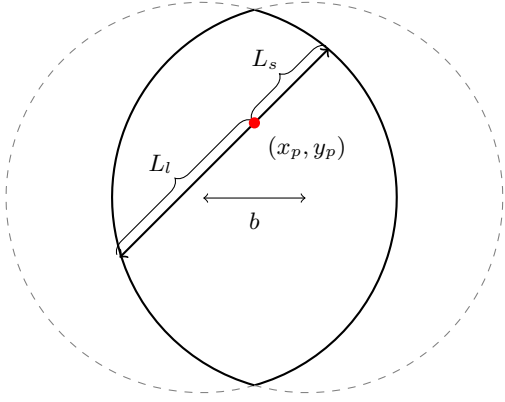
\begin{figure}
\begin{tikzpicture}[scale=0.4]
  \coordinate (C1) at (-1.7,0);
  \coordinate (C2) at (1.7,0);
  \coordinate (P) at (0,2.5);
  \coordinate (I1) at (-4.48112,-1.98112);
  \coordinate (I2) at (2.47226,4.97226);
  \draw[black,<->] (C1) -- node[below=2pt] {$b$} (C2); 
  \draw[black,thick] (0.0,6.26427) arc (74.25:-74.25:6.49084);
  \draw[black,dashed,gray] (0.0,6.26427) arc (74.25:285.74:6.49084);
  \draw[black,thick] (0.0,6.26427) arc (105.75:254.25:6.49084);
  \draw[black,dashed,gray] (0.0,6.26427) arc (105.75:-105.75:6.49084);
  \draw[black,->,thick] (P) -- (I1);
  \draw[decorate,decoration={brace,mirror,raise=0.05cm,amplitude=5pt}] (P) -- node [black,midway,xshift=-10pt,yshift=10pt] {$L_{l}$} (I1);
  \draw[black,->,thick] (P) -- (I2);
  \draw[decorate,decoration={brace,raise=0.05cm,amplitude=5pt}] (P) -- node [black,midway,xshift=-10pt,yshift=10pt] {$L_{s}$} (I2);
  \node[shape=circle,fill=red,inner sep=1.5pt,label=below right:${(x_p,y_p)}$] (X) at (P) {};
\end{tikzpicture}
\caption{Sketch of the initial parton pair traversing the almond-shaped initial collision zone.\label{fig:lengthSketch}}
\end{figure}

For estimating the transverse spatial extension of the initial medium, we use an almond shape resulting from the collision of the two approximately circular nuclei approaching at impact parameter $b$. This picture is schematically sketched in fig.~\ref{fig:lengthSketch}. The radii of these nuclei is estimated by the radius of the employed Woods-Saxon density profile and is parametrized by
\begin{equation}
 R_A = 1.12\,A^{1/3} - 0.86\,A^{-1/3} \,,
\end{equation}
which is e.g. $R_A \approx \SI{6.49}{\fm}$ for \Pb. With that radius and the impact parameter, the two nuclei shapes can be described by circles obeying
\begin{align}
 (x \mp b/2)^2 + y^2 = R_A^2 \, .
\end{align}

Assuming that the partons will fly on straight lines through the medium, which is a good approximation of high energy jets, their trajectory can be described in terms of simple geometry by
\begin{align}
 \begin{pmatrix}x\\y\end{pmatrix} = k \, \begin{pmatrix}p_x^0\\p_y^0\end{pmatrix} + \begin{pmatrix}x_P\\y_P\end{pmatrix} \, ,
\end{align}
where $p_{x}^0 = \frac{p_{x}}{|\vect{p}|}$ and $p_{y}^0 = \frac{p_{y}}{|\vect{p}|}$ are the normalized transverse momentum components and $\left(x_p,y_p\right)$ the spatial creation point of the initial parton pair and $k$ is a proportionality factor.

With that information it is possible to calculate the intersection points between the two nuclei and the parton trajectory and thereby the exit points of the partons from the collision zone. Defining
\begin{equation}
 \Delta x_{L/R} = x_p \pm b/2 \,,
\end{equation}
one can show that the proportionality factor at which the line and the circles intersect, is
\begin{multline}
 k_{L/R} = \pm \sqrt{ \left( \Delta x_{L/R} \, p_x^0 + y_p \, p_y^0 \right)^2 - \Delta x_{L/R}^2 - y_p^2 + R_A^2} \\ - \left( \Delta x_{L/R} \, p_x^0 + y_p \, p_y^0 \right) \, ,
\end{multline}
where $k_{L/R}$ are the respective factors for the intersection with the left ($k_L$) and the right ($k_R$) circle. Because of the previous normalization of the parton momentum vector, the absolute values $|k_{L/R}|$ of the proportionality factors are the in-medium path lengths of the initial parton pair. As a remark, obviously every circle can have two intersection points: Depending on its corresponding $x$-value one has to decide which intersection point is used for determining the respective path length. Having done that, one is able to define the shorter path length as $L_s$ and the longer path length as $L_l$. 

Because of the employed Woods-Saxon density distributions, there is a finite probability for events in which the parton pair is created outside the collision zone. In this case two scenarios are possible: One parton A will fly through the medium and parton B will fly through no medium at all. In this scenario, parton A has path length $L_l \neq \SI{0}{\fm}$ and parton B has path length $L_s = \SI{0}{\fm}$. This will lead to a length imbalance of $L_i = 1$. The other possible scenario is that both partons will not pass the collision zone and thus get path lengths $L_{l/s}=\SI{0}{\fm}$. Obviously, this scenario leads to a length imbalance of $L_i=0$.

\bibliography{mybibliography}

\end{document}